\documentclass[]{iopart}
\usepackage{iopams}
\usepackage{setstack}
\usepackage[]{graphicx}
\usepackage{color}
\usepackage[latin1]{inputenc}
\usepackage{marginnote}

\usepackage{mathbbol}
\usepackage{amssymb}             
\usepackage{textcomp}

\newcommand{\colr}{\color{red}}

\newcommand{\brr}{\bi{r}}
\newcommand{\Real}{{\rm Re}}
\newcommand{\Imag}{{\rm Im}}

\newcommand{\we}{\mathbb{w}}

\newcommand{\cA}{{\mathcal{D}}}
\newcommand{\tw}{{w'}}

\begin{document}

\title[LISA's wavefront errors]{Coupling of wavefront errors and pointing jitter in the LISA interferometer:
misalignments of the interfering wavefronts}
\author{C P Sasso$^1$, G Mana$^1$, and S Mottini$^2$}
\address{$^1$INRIM -- Istituto Nazionale di Ricerca Metrologica, Str. delle cacce 91, 10135 Torino, Italy}
\address{$^2$Thales Alenia Space, Str. Antica di Collegno, 253, 10146 Torino, Italy}
\ead{c.sasso@inrim.it}

\begin{abstract}
The Laser Interferometer Space Antenna is a foreseen space-based gravitational wave detector, which aims to detect $10^{-20}$ strains in the frequency range from 0.1 mHz to 1 Hz. It is a triangular constellation of three spacecraft, with equal sides of $2.5 \times 10^9$ m, where every spacecraft hosts a pair of telescopes that simultaneously transmit and receive laser beams measuring the constellation arms by heterodyning the received wavefronts with local references. Due to the spacecraft and constellation jitters, the interfering (received and local) wavefronts become misaligned. We investigate analytically the coupling between misalignments and aberrations of the interfering wavefronts and estimate the relevant contribution to the noise of the heterodyne signal.
\end{abstract}

\submitto{Classical and Quantum Gravity}

\pacs{42.15.Dp, 07.60.Ly, 04.80.Nn, 95.55.Ym}

\section{Introduction}
The spacecraft of the Laser Interferometer Space Antenna (LISA) are at the vertices of an equilateral triangle, which is in a plane inclined 60$^\circ$ with respect to the ecliptic and which trails behind the Earth by 20$^\circ$ \cite{CDF:2017,GWOAdvisory:2016,Danzmann:2017}. Each spacecraft is equipped with two telescopes that simultaneously transmit and receive 1064 nm laser beams linking the constellation by heterodyning the received wavefronts with local references. {\colr With a 135$\times$ magnification and a diameter of the primary mirror of 300 mm, the received beam to the heterodyne detection has a diameter of 2.2 mm \cite{ESA:2017}.} A critical aspect of LISA is the sensitivity required in the measurement of the spacecraft separations, $2.5 \times 10^9$ m: the measurement noise must approach 1 pm/$\sqrt{{\rm Hz}}$ in the frequency band from 0.1 mHz to 1 Hz \cite{Armano:2016}. This requires that the noise of the interferometric phase measurement is near to 1 $\mu$rad/$\sqrt{{\rm Hz }}$, and imposes tight requirements on the stability of the interfering beams.

{\colr  Aberrations and jitter of the wavefront sent by a spacecraft to the next cause a measurement noise. In a previous paper, we investigated the propagation of the wavefront aberrations and the noise of the measured distance originated by the coupling of the aberrations with the transmitter jitter \cite{Sasso:2018}. We found that the sensitivity to the jitter increases from $0.02$ pm/nrad to $0.28$ pm/nrad when the optical quality of the transmitted wavefront decreases from $\lambda/40$ to $\lambda/10$. In this paper, to complete the noise assessment, we investigate the phase noise originated by the coupling of an aberrated reception with the receiver jitter.}

Owing to the spacecraft jitter and the breathing of the constellation, the received wavefront -- demagnified, truncated, and aberrated -- is misaligned with respect to the local reference and -- in the presence of a non-uniform phase profile of the interference pattern -- this causes the measurement of an apparent distance in excess (or defect) to the actual one. {\colr The requirement is that the coupling between tilt and measured distance is less than 25 pm/$\mu$rad for misalignments of the interfering beams to within $\pm 300$ $\mu$rad \cite{Chwalla:2016}}

Experimental investigations of the tilt to length coupling were carried by using imaging systems and test beds representative of the LISA's operation \cite{Schuster:2016,Chwalla:2016}. Our paper analytically describes and predicts the combined effect of misalignments and aberrations of the interfering wavefronts -- expressed in terms of the Zernike modal amplitudes -- on the phase of the heterodyne signal. It gives the phase in excess (or defect) for the interference of beams having truncated Gaussian profile. Eventually, it takes advantage of this result to carry out a Monte Carlo simulation of the interferometer operation and to develop criteria for the assessment of the phase noise.

\section{Heterodyne interferometry}
We describe the local reference, $E_1(\brr;t)=u_1(\brr)$, and received, $E_2(\brr;t)=u_2(\brr)\rme^{\rmi\Omega t}$, optical fields on the detector plane by the complex amplitudes
\numparts\begin{equation}\label{u1}
 u_1(\brr) = \rme^{-r^2/w_1^2} \rme^{-\rmi\we_1(\brr)} ,
\end{equation}
end
\begin{equation}\label{u2}
 u_2(\brr) = \rme^{-r^2/w_2^2} \rme^{-\rmi\we_2(\brr)} ,
\end{equation}\endnumparts
where $\we_1(\brr)$ and $\we_2(\brr)$ are small, zero mean, deviations from flat wavefronts, $\brr=(x,y)^T$ is a position vector in the detector plane, and $\Omega$ is the heterodyne angular frequency. We did not consider more realistic intensity profiles of the received beam, assuming Gaussian profiles, having 1/e$^2$ radii equal to $w_1$ and $w_2$.

In (\ref{u2}), we omitted the unessential phase retardation related to the spacecraft separation, $-\rmi kz$, where $k=2\pi/\lambda$ is the wave vector, and $z$ is the spacecraft distance, and the common term $\rme^{\rmi \omega t}$. Furthermore, for the sake of simplicity, we assume an imaging system fixing any wandering of the received beam across the detector \cite{Schuster:2016}. By setting to zero the piston terms of $\we_1(\brr)$ and $\we_2(\brr)$, we do not consider the phase retardation related to the optical lengths of the beam paths through the telescope and optical bench. 

The heterodyne signal is given by alternating part of
\numparts\begin{equation}
 S = \int_\cA |E_1(\brr) + E_2(\brr)|^2 \,\rmd\brr =
 G \left[ 1 + \Gamma \cos(\Omega t + \phi) \right] ,
\end{equation}
where
\begin{equation}\label{G}
  G = \int_\cA \left( |u_1(\brr)|^2 + |u_2(\brr)|^2 \right) \,\rmd\brr
\end{equation}
is the direct signal,
\begin{equation}\label{visibility}
  \Gamma = \frac{2|\Xi|}{G}
\end{equation}
is the signal visibility,
\begin{equation}\label{fase}
  \phi = \arg(\Xi)
\end{equation}
is the phase of the heterodyne signal in excess (or in defect) to the $-kz$ retardation,
\begin{equation}
  \Xi = \int_\cA u_1^*(\brr)u_2(\brr) \,\rmd \brr ,\label{Xi}
\end{equation}\endnumparts
is the complex amplitude of the alternating signal, and $\cA$ is the detector area -- a disk having $r_0\approx 2.5$ mm radius. Therefore, the extra phase is
\begin{equation}\label{phase}
 \phi = \arg \left[ \int_0^{r_0} r \rme^{-2r^2/w^2} \left( \int_0^{2\pi} \rme^{\rmi\we(\brr)}\, \rmd\theta \right)\,\rmd r \right] ,
\end{equation}
where the $w^2=2w_1^2 w_2^2/(w_1^2+w_2^2)$ is the harmonic mean of $w_1^2$ and $w_2^2$, $\we(\brr)=\we_2(\brr)-\we_1(\brr)$ is the deviation from flatness of the interference-pattern phase, and $r$ and $\theta$ are the radial and azimuthal coordinates. It is worth noting that only the differential error contributes to the error budget, not the common-mode errors of each wavefront. In the case of identical and perfectly overlapped wavefronts, that is, if $\we(\brr)=0$, the detection of the received-wavefront phase is free of errors.
\section{Tilt to aberrations coupling}
\subsection{Wavefront misalignment.}
A misalignment between the interfering wavefronts transforms the phase profile of the interference pattern as
\begin{equation}\label{tilt}
 \we(\brr) \rightarrow \we(\brr) + k \brr\cdot\balpha = \we(r,\theta) + kr \alpha\cos(\theta-\beta) ,
\end{equation}
where $\balpha=[\alpha_x, \alpha_y]^T=\alpha[\cos(\beta), \sin(\beta)]^T$ is a small rotation about an axis lying in the detector plane and having azimuth $\beta$, $\brr=[x, y]^T=r[\cos(\theta), \sin(\theta)]^T$, the pivot -- the piston term being omitted -- is in the origin of the reference frame, and we considered only the first order term.

If the only aberration of the interference pattern is a tilt, that is, if the errors otherwise the misalignment of the interfering wavefronts match perfectly, by using (\ref{tilt}) in (\ref{Xi}), we obtain
\begin{equation}\fl\label{Xi0}
 \Xi = \int_0^{r_0} r\, \rme^{-2r^2/w^2} \,\rmd r \int_0^{2\pi} \rme^{\rmi kr\alpha\cos(\theta-\beta))}\, \rmd\theta =
 2\pi \int_0^{r_0} r\, \rme^{-2r^2/w^2} {\rm J}_0(kr\alpha)\,\rmd r ,
\end{equation}
where ${\rm J}_0(\Box)$ is the Bessel function of the first kind of order zero, and, consequently, $\phi=0$. Therefore, as noted in \cite{Schuster:2015}, apart from a contrast loss, the interference of identical wavefronts is insensitive to misalignments. It is worth noting that this is true also when the intensity radial-profiles deviate from Gaussian ones and the detector radius is finite.

\subsection{Zernike modal amplitudes.}
In the following, we will use the approximation
\begin{equation}\label{w-app}
 \rme^{\rmi \we(\brr)} \approx 1 + \rmi\we(\brr) - \frac{1}{2} \we^2(\brr) - \frac{\rmi}{6} \we^3(\brr) + ...
\end{equation}
and express the phase profile of the interference pattern in terms of the Zernike modal amplitudes. Therefore,
\begin{equation}\label{we0}
 \we(\brr) = \sum_{n=1}^\infty \sum_{m=-n}^n z_n^m R_n^{|m|}(\rho) \rme^{\rmi m\theta} ,
\end{equation}
where $R_n^{|m|}(\rho)=0$ for all $n-|m|$ odd or negative, $\rho=|\brr|/r_0$ is the normalized radial coordinate, $\theta$ is the azimuth, and the radial polynomials satisfy the orthogonality relation
\begin{equation}\label{Z-ortho}
 \int_0^1  R_n^{|m|}(\rho) R_{n'}^{|m|}(\rho) \rho\,\rmd \rho = \frac{\delta_{n,n'}R_n^{|m|}(1)}{2(n+1)} .
\end{equation}
It must be noted that, following (\ref{u1}-$b$), the Zernike modal amplitudes $z_n^m$ are expressed in radians.

The relationship $z_n^{-m}=z_n^{m*}$ ensures that (\ref{we0}) is real. Hence, $z_n^0$ is real and, if $m\ne 0$,
\begin{equation}
 z_n^{\pm m} = |z_n^m| \rme^{\pm\rmi\theta_n^m}
\end{equation}
Eventually, the $n=1$ term of (\ref{we0}),
\begin{equation}\label{z11}
 z_1^{-1} \rho \rme^{-\rmi\theta} + z_1^1 \rho \rme^{\rmi\theta} = 2|z_1^1| \rho \cos(\theta+\theta_1^1) ,
\end{equation}
takes the misalignment of the interfering wavefronts,
\begin{equation}\label{alpha}
 \alpha = \frac{2}{kr_0} |z_1^1| ,
\end{equation}
into account.

\subsection{Phase of the heterodyne signal.}
{\colr The lowest order coupling between wavefront misalignments and aberrations entails products of three Zernike modal amplitudes. Therefore, by using the series expansion (\ref{w-app}) up to the third order} and measuring the radius of the interference pattern $w$ in $r_0$ units, we write the alternating signal (\ref{Xi}) as
\numparts\begin{eqnarray}\nonumber
 \frac{2\Xi}{\pi \tw^2} &=& \frac{2}{\pi\tw^2} \int_0^1 \rho\, \rme^{-2\rho^2/\tw^2} \,\rmd\rho
 \int_0^{2\pi} \rme^{\rmi\we(\rho,\theta)} \,\rmd\theta \\ \label{Xi-app}
      &=& a_0 + \rmi a_1 + a_2 + \rmi a_3 ,
\end{eqnarray}
where $\tw=w/r_0$ and
\begin{eqnarray}\label{a1}
 a_0 &=& \frac{4}{\tw^2} \int_0^1 \rho\, \rme^{-2\rho^2/\tw^2} \,\rmd\rho = 1 - \rme^{-2/\tw^2} , \\ \label{a234}
 a_1 &=& \frac{2}{\pi\tw^2} \int_0^1 \rho\, \rme^{-2\rho^2/\tw^2} \,\rmd\rho \int_0^{2\pi} \we(\rho,\theta) \, \rmd\theta , \\
 a_2 &=& -\frac{1}{\pi\tw^2}\int_0^1 \rho\, \rme^{-2\rho^2/\tw^2} \,\rmd\rho \int_0^{2\pi} \we^2(\rho,\theta)\, \rmd\theta , \\
 a_3 &=& -\frac{1}{3\pi\tw^2} \int_0^1 \rho\, \rme^{-2\rho^2/\tw^2} \,\rmd\rho \int_0^{2\pi}  \we^3(\rho,\theta)\, \rmd\theta .
\end{eqnarray}\endnumparts
Apart from the factor of two in the exponential and an unessential scale factor, (\ref{Xi-app}) is identical to the third-order approximation of the Rayleigh-Sommerfeld integral giving the on-axis far-field of a truncated aberrated wavefront having a Gaussian intensity profile \cite{Sasso:2018}. {\colr We carried out the integrations (\ref{a234}-$e$) analytically with the aid of Mathematica \cite{Mathematica}. We do not give the results here, but the code is available in the supplementary material.}

By approximating $\arg(\Xi)$ as $\Imag(\Xi)/\Real(\Xi)$, limiting the wavefront aberration to tilt, defocus, astigmatism, coma, trefoil, and spherical (i.e., by considering only the modal amplitudes $z_1^1, z_2^0, z_2^2, z_3^1, z_3^3$, and $z_4^0$), {\colr using the results of the (\ref{a234}-$e$) integrations, and considering only the lowest order terms}, the extra phase of the heterodyne signal (\ref{fase}) is
\begin{eqnarray}\nonumber
 \phi &\approx \frac{ (a_1+a_3) ( 1 - \rme^{-2/w'^2} - a_2 ) }{(1 - \rme^{-2/w'^2})^2 } \\ \label{phi}
      &\approx b_{00} + b_{10} \zeta_x + b_{20} \zeta_x^2 + b_{01} \zeta_y + b_{02} \zeta_y^2 + b_{11} \zeta_x\zeta_y .
\end{eqnarray}
In (\ref{phi}), we made explicit the dependence on the horizontal and vertical tilt aberrations,
\numparts\begin{eqnarray}\label{hvt}
 \zeta_x &= &|z_1^1|\cos(\theta_1^1) = kr_0\alpha_x/2 , \\
 \zeta_y &= &|z_1^1|\sin(\theta_1^1) = kr_0\alpha_y/2 ,
\end{eqnarray}\endnumparts
where $\alpha_x$ and $\alpha_y$ are the horizontal and vertical components of the wavefront misalignment $\alpha$. When calculating the $b_{ij}$ coefficients, which are given in the appendix, we carried out the reparametrization from the $|z_1^1|$ and $\theta_1^1$ pair to the $\zeta_x$ and $\zeta_y$ one with the aid of Mathematica \cite{Mathematica}. The code is available in the supplementary material.

The parabolic approximation (\ref{phi}) considers the aberration contribution up to the third order. It requires that $|\we(\brr)| < 1$, which -- according to (\ref{z11}) -- implies $2|z_1^1| <1$. Therefore, by using $\lambda=1064$ nm and $r_0 \approx 1.1$ mm in (\ref{alpha}), the equation (\ref{phi}) is valid if $\alpha < 155$ $\mu$rad. The validity of the approximation (\ref{phi}) for larger misalignments will be examined numerically in section \ref{results}.

\subsection{Visiblity of the heterodyne signal.}
\begin{figure}
\includegraphics[width=6.5cm]{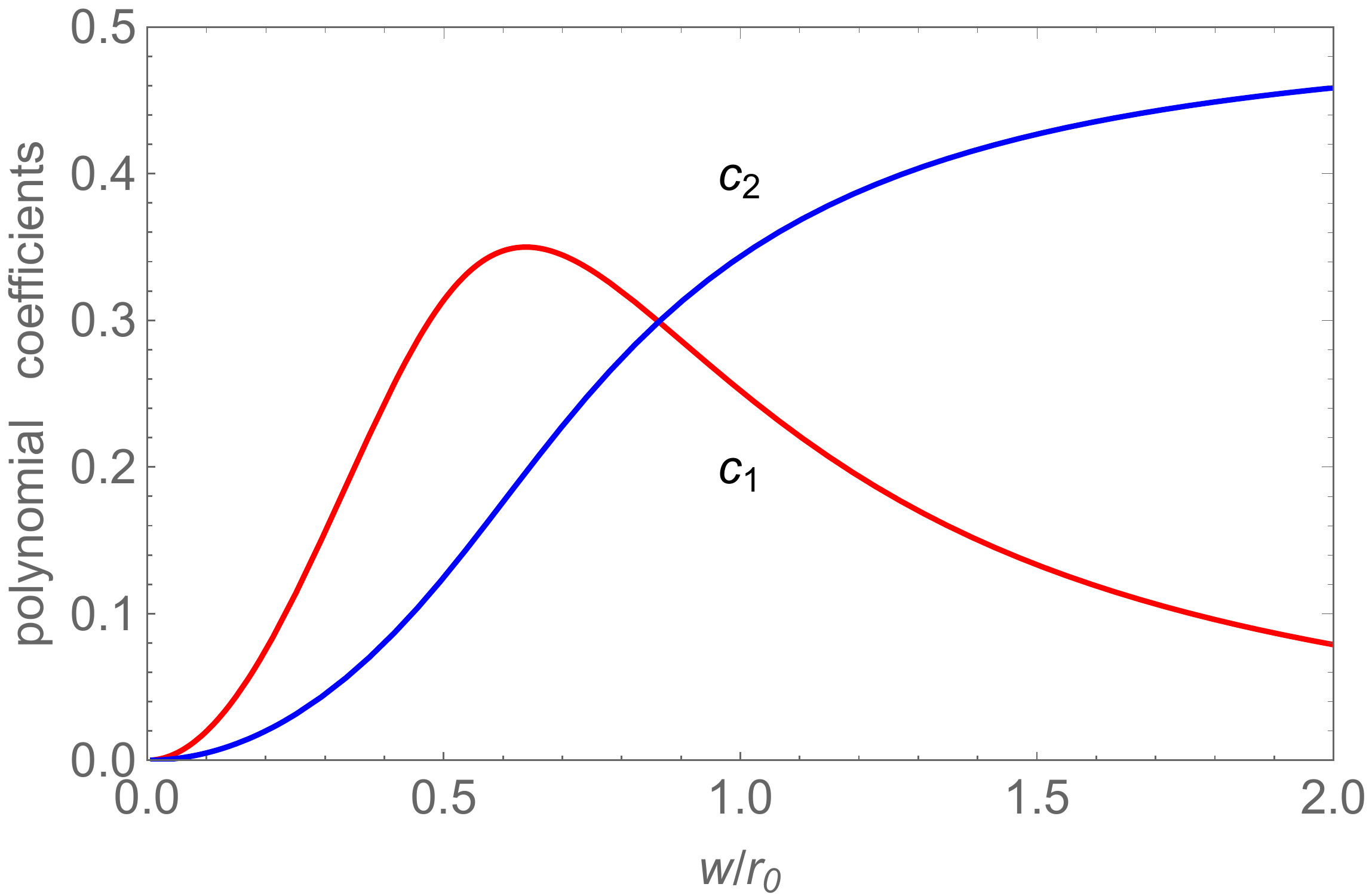}
\caption{Coefficients of the polynomial (\ref{gamma}).}\label{Fig-c12}
\vspace{2mm}
\includegraphics[width=6.25cm]{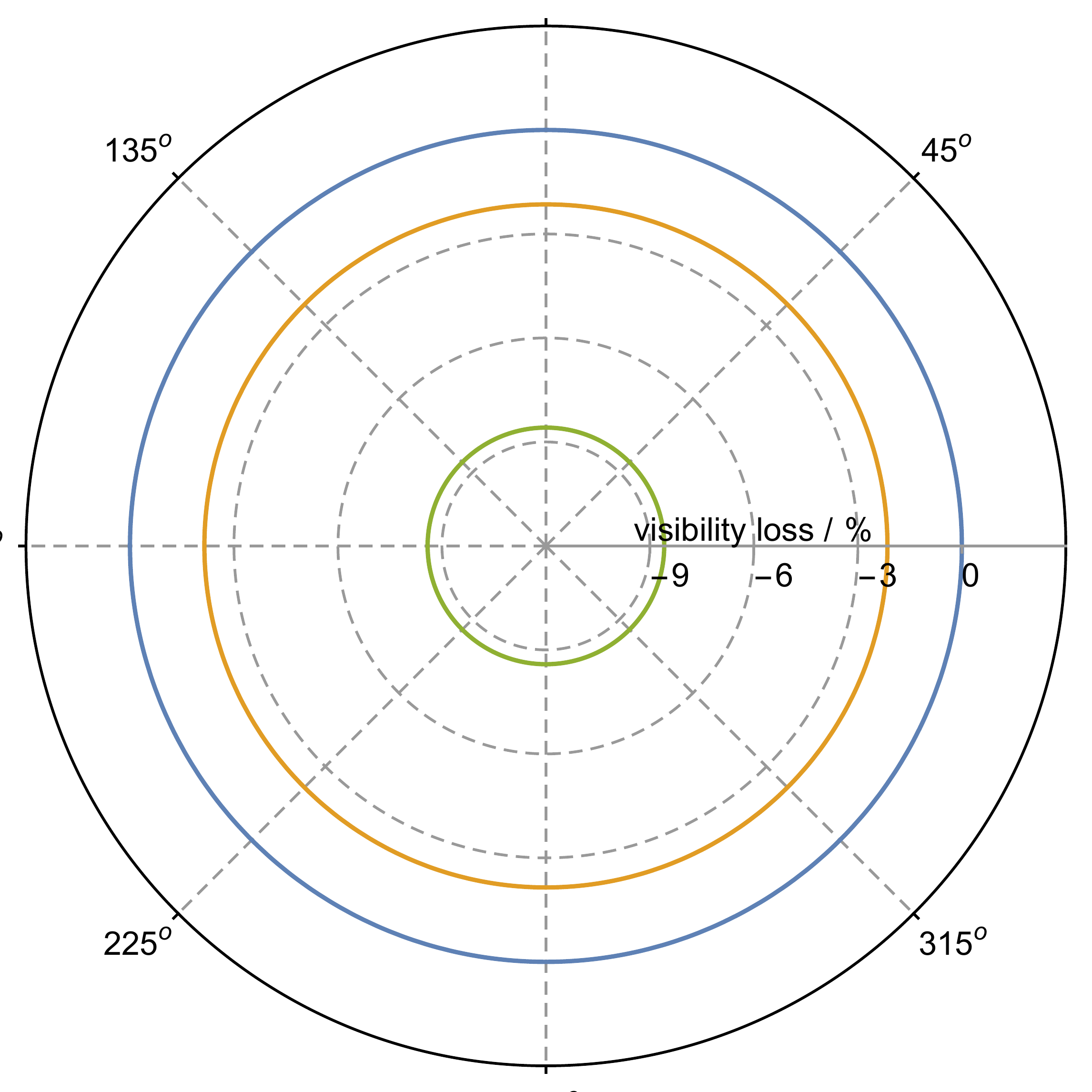}
\includegraphics[width=6.25cm]{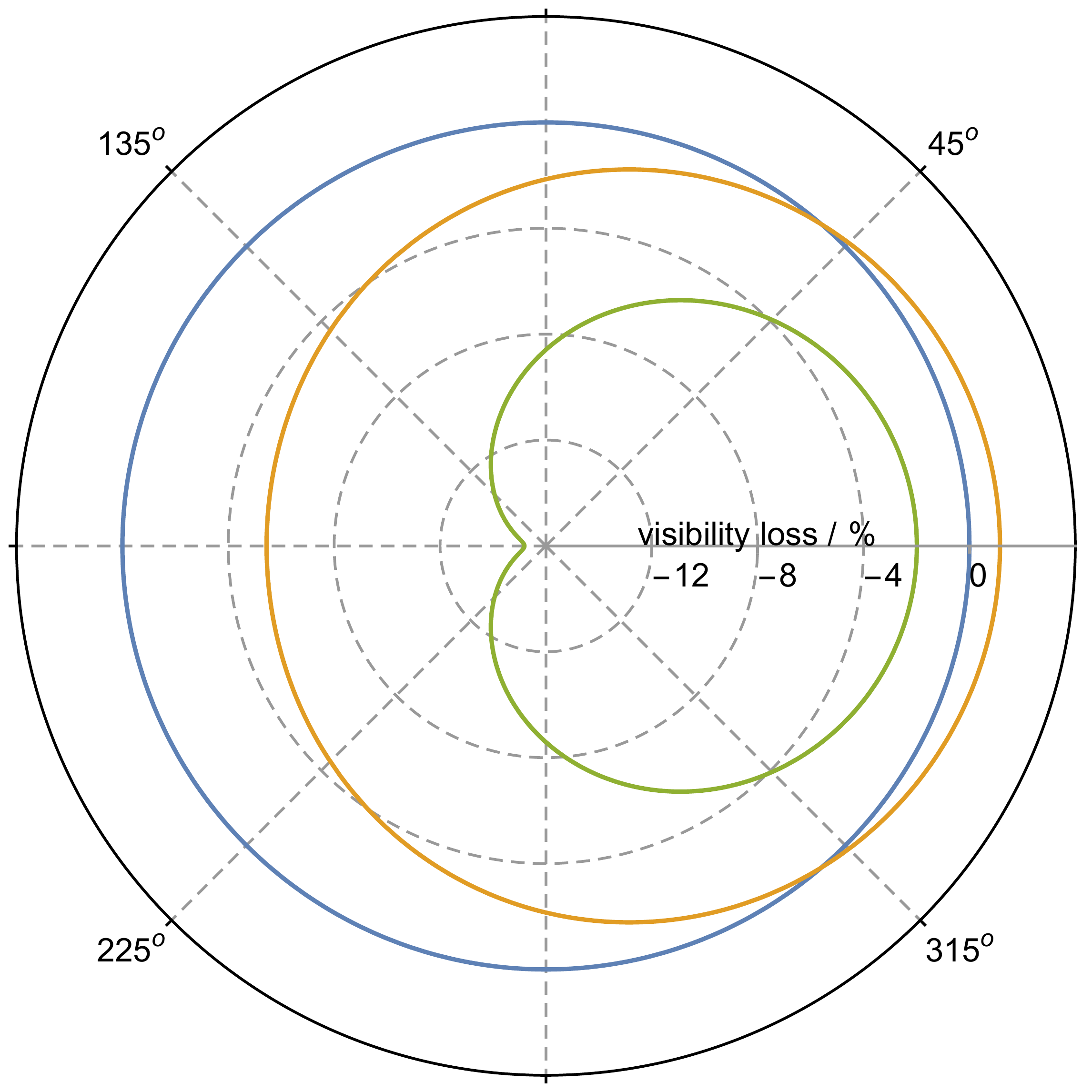}
\caption{Polar plot of the visibility loss/gain of the heterodyne signal. The angular coordinate is the $z_1^1$ azimuth relative to the $z_3^1$ one; $w/r_0$ has been set to one. Blue: $|z_1^1|=0$ rad, orange: $|z_1^1|=0.25$ rad (if $r_0=1.1$ mm, 77 $\mu$rad misalignment), green: $|z_1^1|=0.5$ rad (if $r_0=1.1$ mm, 154 $\mu$rad misalignment). Left: $|z_3^1|=0$ rad. Right: $|z_3^1|=0.52$ rad.}\label{Fig-visibility}
\vspace{2mm}
\includegraphics[width=6.5cm]{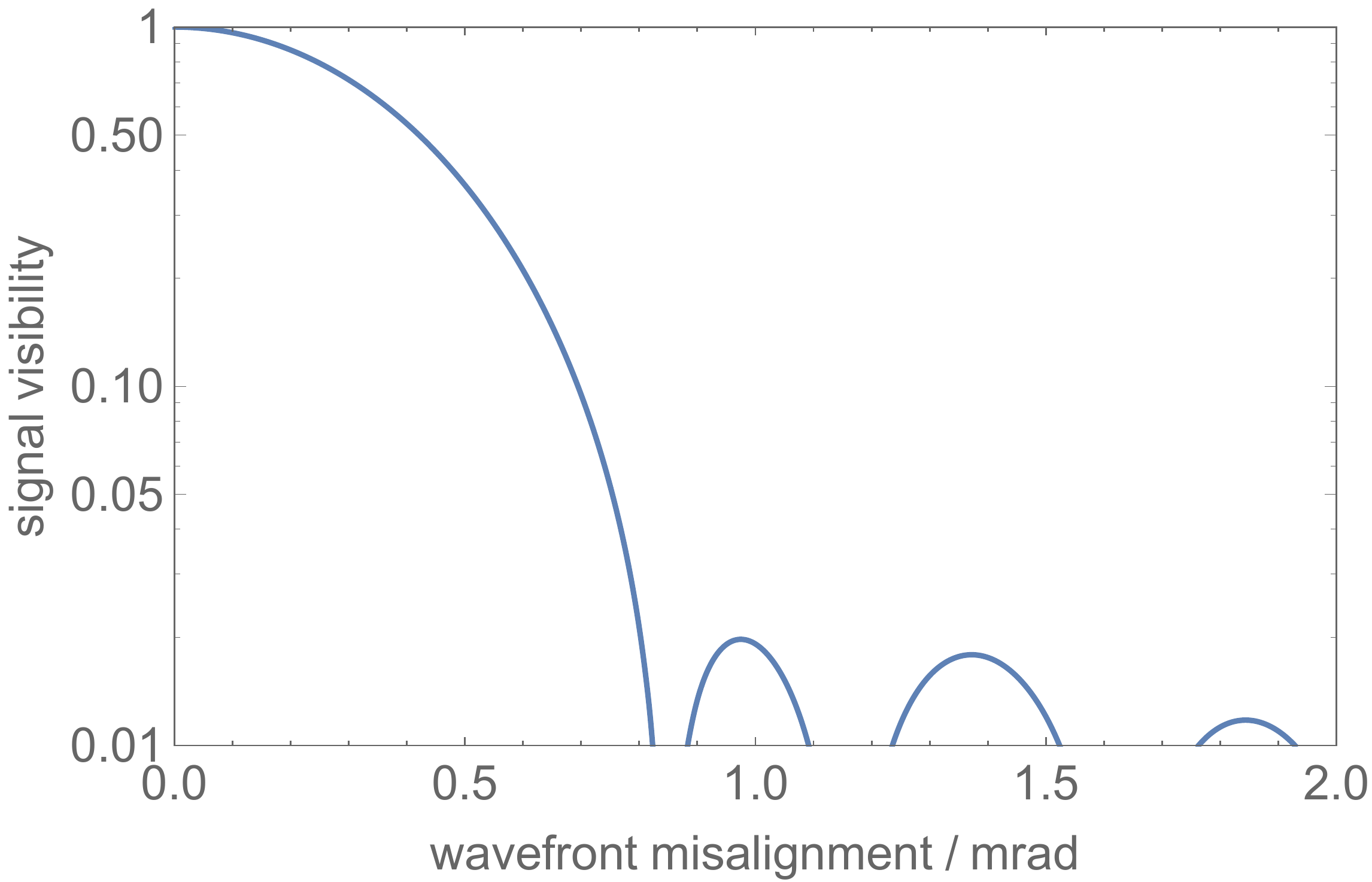}
\caption{Visibility of the heterodyne signal for large wavefront misalignments when the wavefront aberrations match and $w/r_0=1$.}\label{Fig-visibility3}
\end{figure}

By remembering (\ref{visibility}) and (\ref{Xi-app}) and up to second order of the Zernike modal amplitudes, the visibility of the heterodyne signal is
\begin{eqnarray}\nonumber
 \Gamma   &\approx &\frac{\pi \tw^2\big| a_0 + \rmi a_1 + a_2 \big|}{G} \\ \nonumber
 &\approx &\frac{\sqrt{ a_0^2 + 2a_0 a_2 + a_1^2 }}{a_0} \\ \label{gamma-0}
 &=       &\frac{\sqrt{ (1 - \rme^{-2/w'^2})^2 + 2(1 - \rme^{-2/w'^2})a_2 + a_1^2 }}{1 - \rme^{-2/w'^2}} ,
\end{eqnarray}
where, for the sake of simplicity, we assumed $w_1=w_2=w'$ and used $G=\pi \tw^2 a_0$. Next, {\colr by using (\ref{a234}-$d$) and making the dependence on $z_1^1$ explicit}, the signal visibility is
\numparts\begin{equation}\label{gamma}
 \Gamma \approx 1 + c_1 |z_3^1| \cos(\theta) |z_1^1| - c_2 |z_1^1|^2 ,
\end{equation}
where $\theta=\theta_1^1-\theta_3^1$ is the tilt-aberration azimuth relative to the one of the coma,
\begin{eqnarray}
  c_1 &=& - \frac{2 + 2 (2 + \rme^{2/\tw}) \tw^2 + 3 (1 - \rme^{2/\tw}) \tw^4)}{1-\rme^{2/\tw}} ,\\
  c_2 &=&  \frac{2 + (1 - \rme^{2/\tw}) \tw^2}{2(1-\rme^{2/\tw})} ,
\end{eqnarray}\endnumparts
and we considered only the lowest-order terms. The Figure \ref{Fig-c12} shows how the $c_{1,2}$ coefficients depends on the interference-pattern truncation. {\colr The symbolical calculations implied in obtaining (\ref{gamma}-$c$) from (\ref{gamma-0}) where carried out with the aid of Mathematica \cite{Mathematica}; the code is available in the supplementary material.}

Figure \ref{Fig-visibility} shows the impacts of the coupling between misalignments and aberrations of the interfering wavefronts on the signal visibility when $w/r_0=1$. The left part shows that, when the coma is null, the visibility decreases quadratically as the wavefront misalignment increases. As shown in Fig.\ \ref{Fig-visibility} (right), a wavefront misalignment at the same azimuthal angle of the coma increases the signal visibility. Conversely, a misalignment at the opposite azimuthal angle reduces the visibility. These gain and loss of visibility occur because of the beam's perception -- {\colr which depends on the ratio between the radii of the interference-pattern and detector} -- of a misalignment associated to the coma.

In the case of large misalignments of the interfering wavefronts, the wavefront aberrations have a negligible effect on the signal visibility and can be neglected. Hence, by assuming again $w_1=w_2=w$ and using (\ref{Xi0}) and
\begin{equation}
 G = (1-\rme^{-2/w'^2})\pi w'^2
\end{equation}
in (\ref{visibility}), we obtain
\begin{equation}
 \Gamma = \frac{4\displaystyle{\int_0^1} \rho\, \rme^{-2\rho^2/w'^2} {\rm J}_0(kr_0\alpha\rho)\,\rmd \rho }{(1-\rme^{-2/w'^2})w'^2} ,
\end{equation}
which is shown in Fig.\ \ref{Fig-visibility3} when $w/r_0 = 1$. For a wavefront misalignment of 300 $\mu$rad, the visibility reduces to 71\%.

\section{Phase noise}
The pointing jitter of the receiving telescope translates in a (magnified) jitter of the interfering wavefronts. {\colr Therefore, by linearization of (\ref{phi}) and propagation of the tilt-aberration noise, the variance of the heterodyne-signal phase is}

\begin{equation}\fl\label{noise-var}
 \sigma_\phi^2 \approx (b_{10} + 2b_{20} \zeta_{0x} + b_{11} \zeta_{0y})^2 \sigma_{\zeta x}^2 +
 (b_{01} + 2b_{02} \zeta_{0y} + b_{11} \zeta_{0x})^2 \sigma_{\zeta y}^2 ,
\end{equation}

where $\zeta_{0x}$ and $\zeta_{0y}$ are the mean (\ref{hvt}-$b$) tilts and $\sigma_{\zeta x}^2$ and $\sigma_{\zeta y}^2$ are the variances of the zero-mean and uncorrelated jitters of the tilt aberration.

Actually, (\ref{noise-var}) is not very useful because it depends on the Zernike modal amplitudes of $\we(x,y)$. To find a criterion for the assessment of the wavefront quality, we decided to use an expression determined by the $\we(x,y)$ flatness alone. {\colr This requires the averaging of (\ref{we0}) with respect to $z_n^m$ (excluded $z_1^1$) constrained to a predetermined flatness of $\we(x,y)$. After the averaging, by assuming the $\theta_n^m$ angles uniform in the $[0,2\pi]$ interval and observing how the $b_{ij}$ coefficients given in (\ref{bxy}-$f$) depend on these angles, (\ref{noise-var}) does not depend any more on the first powers and product of $\zeta_x$ and $\zeta_y$. In fact, only a constant and terms proportional to $\zeta_{0x}^2$ and $\zeta_{0y}^2$ withstand the average. Therefore, in the simplest case where the jitter is isotropic -- that is, $\sigma_{\zeta x} = \sigma_{\zeta y} = kr_0 \sigma_0/2$ -- we can write the average variance as
\begin{equation}\label{mean-phi}
 \langle \sigma_\phi^2 \rangle_{z_n^m} \approx (g_0 + g_2\alpha_0^2) \sigma_\alpha^2 ,
\end{equation}
where $\langle \square \rangle_{\tau}$ is the average of $\square$ calculated with respect the distribution of $\tau$, we substituted the $\alpha_x$ and $\alpha_y$ misalignments for the $\zeta_x$ and $\zeta_y$ tilts by using (\ref{hvt}-$b$), $\alpha_0 = (\alpha_{0x}^2 + \alpha_{0y}^2)^{1/2}$ is the misalignment, $\sigma_\alpha^2 = 4\sigma_0^2$ is the jitter variance, and the $g_0$ and $g_2$ coefficients -- which depend on the $\we(x,y)$ flatness -- must be calculated numerically.}

\section{Results}\label{results}
To give a calculation example, {\colr in the absence of information on the $\we(x,y)$ spectrum, in the Monte Carlo simulation the trigonometric components of the Zernike modal amplitudes -- $z_2^0, z_2^2, z_3^1$, $z_3^3$, and $z_4^0$ -- were drawn from less-informative, identical, and zero-mean Gaussian distributions constrained to a deviation of $\we(x,y)$ from flatness equal to $\lambda/20$. We imposed this constraint by calculating the deviation -- say, $\lambda/p$ -- and by re-scaling the picked amplitudes by the $p/20$ ratio.} The detector radius and beam wavelength were set to $r_0=1.1$ mm and $\lambda=1064$ nm, respectively. The normalized radius $w'=w/r_0$ of the interference pattern was set to one. {\colr Next, for each Monte Carlo run, the $b_{ij}$ coefficients (\ref{bxy}) were calculated and stored for the subsequent calculation of (\ref{phi}) and (\ref{noise-var}).}

\begin{figure}
\includegraphics[width=7.5cm]{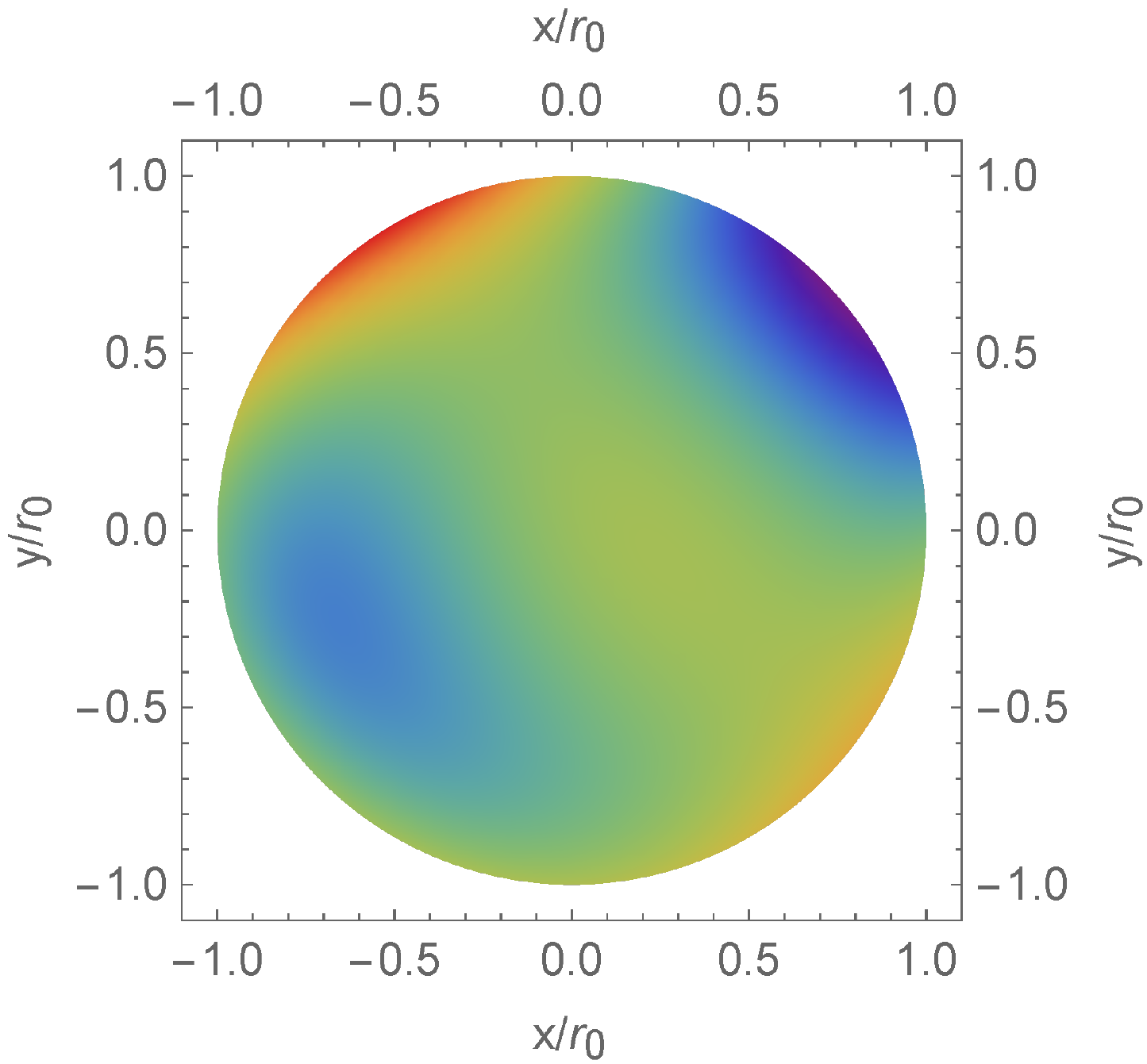}
\caption{Phase profile of the interference pattern. The colour scale is from $-26.6$ nm (violet) to $+26.6$ nm (red); $x$ and $y$ are the transverse coordinates in the detection plane; $r_0=1.1$ mm is the detector radius.}\label{Fig-phase-profile}
\vspace{5mm}
\includegraphics[width=6.5cm]{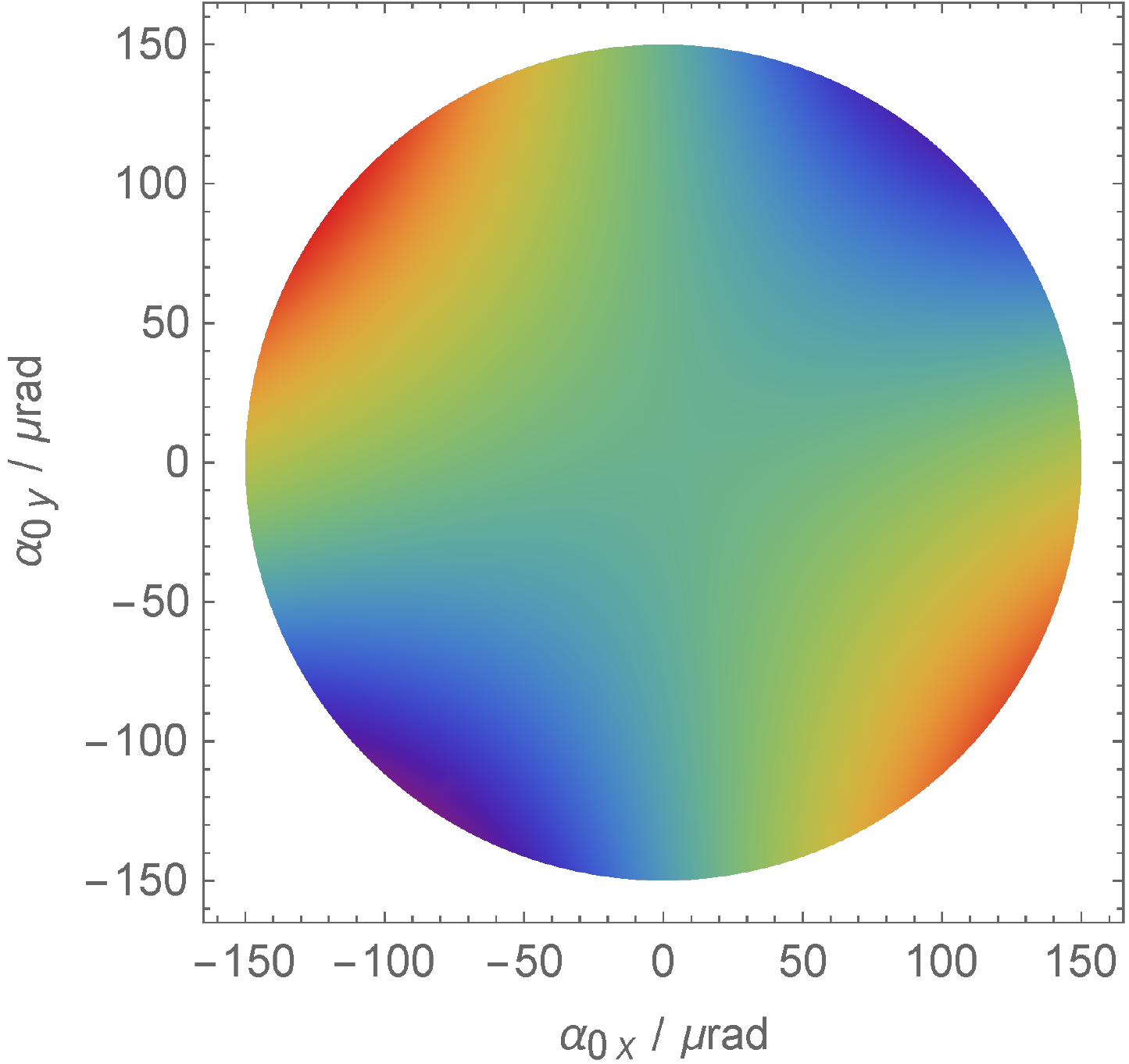}
\includegraphics[width=6.5cm]{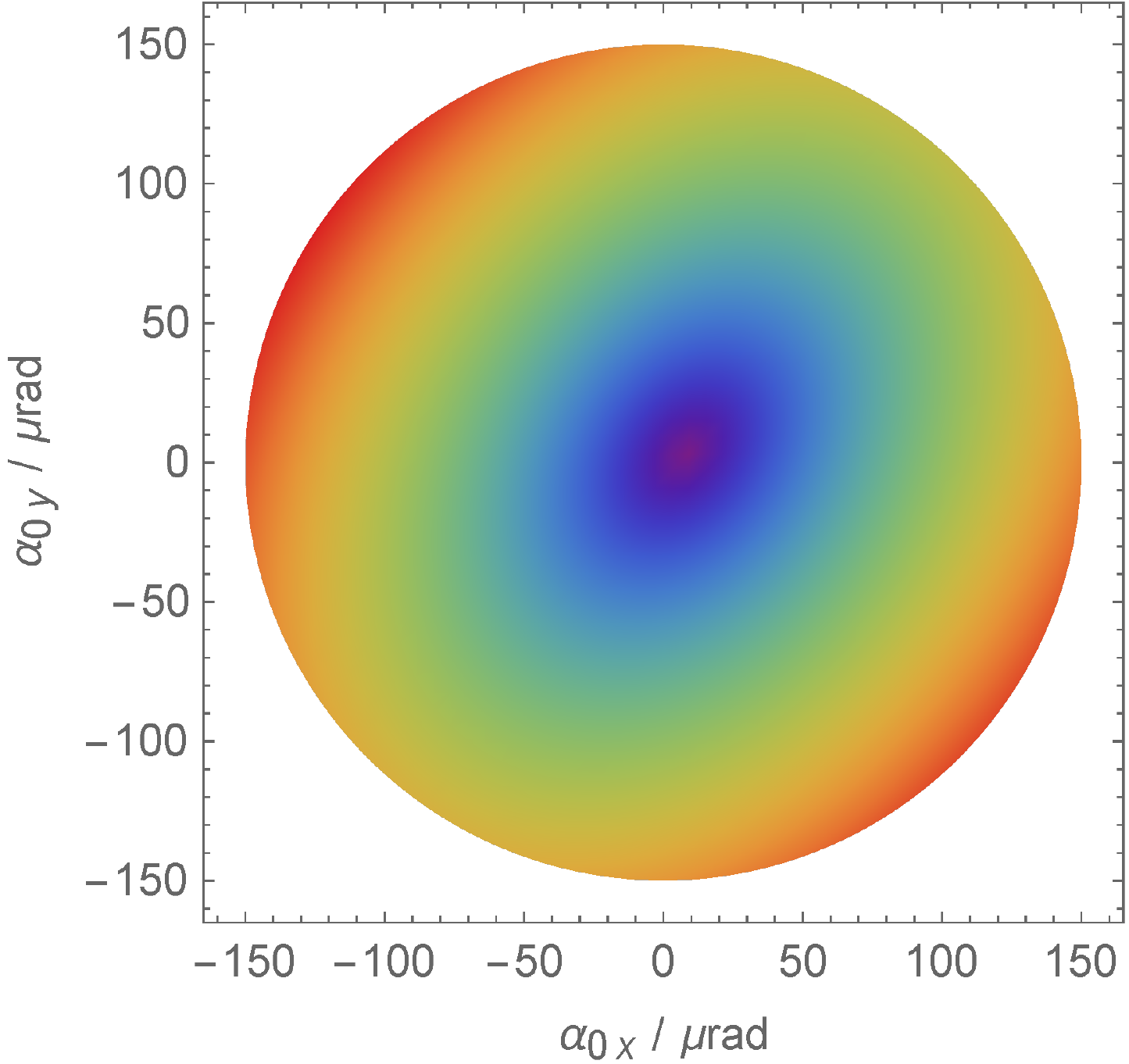}
\caption{Left: phase of the heterodyne signal (\ref{phi}). The colour scale is from zero (violet) to $1.5$ nm (red). Right: standard deviation of the phase noise of the heterodyne signal due to 100 nrad (standard deviation) jitters of the horizontal and vertical wavefront misalignments. The colour scale is from zero (violet) to $1.1$ nm (red). $\alpha_{0x}$ and $\alpha_{0y}$ are the mean misalignments. The detector radius is $r_0=1.1$ mm, the normalized radius of the interference pattern is $w/r_0=1$. The phase profile of the interference pattern is shown in Fig.\ \ref{Fig-phase-profile}.}\label{Fig-h-signal}
\end{figure}

\begin{figure}
\includegraphics[width=6.5cm]{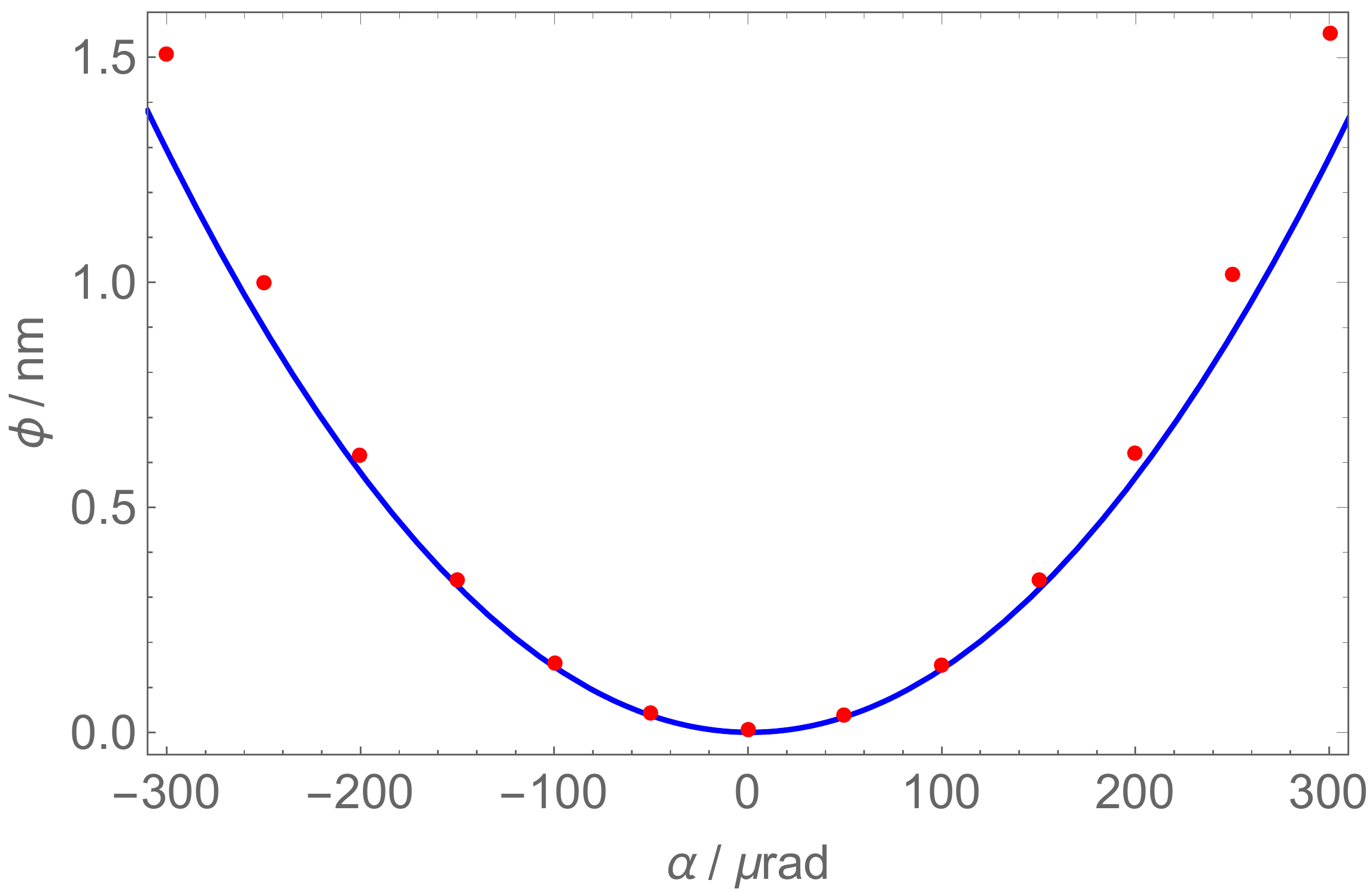}
\includegraphics[width=6.5cm]{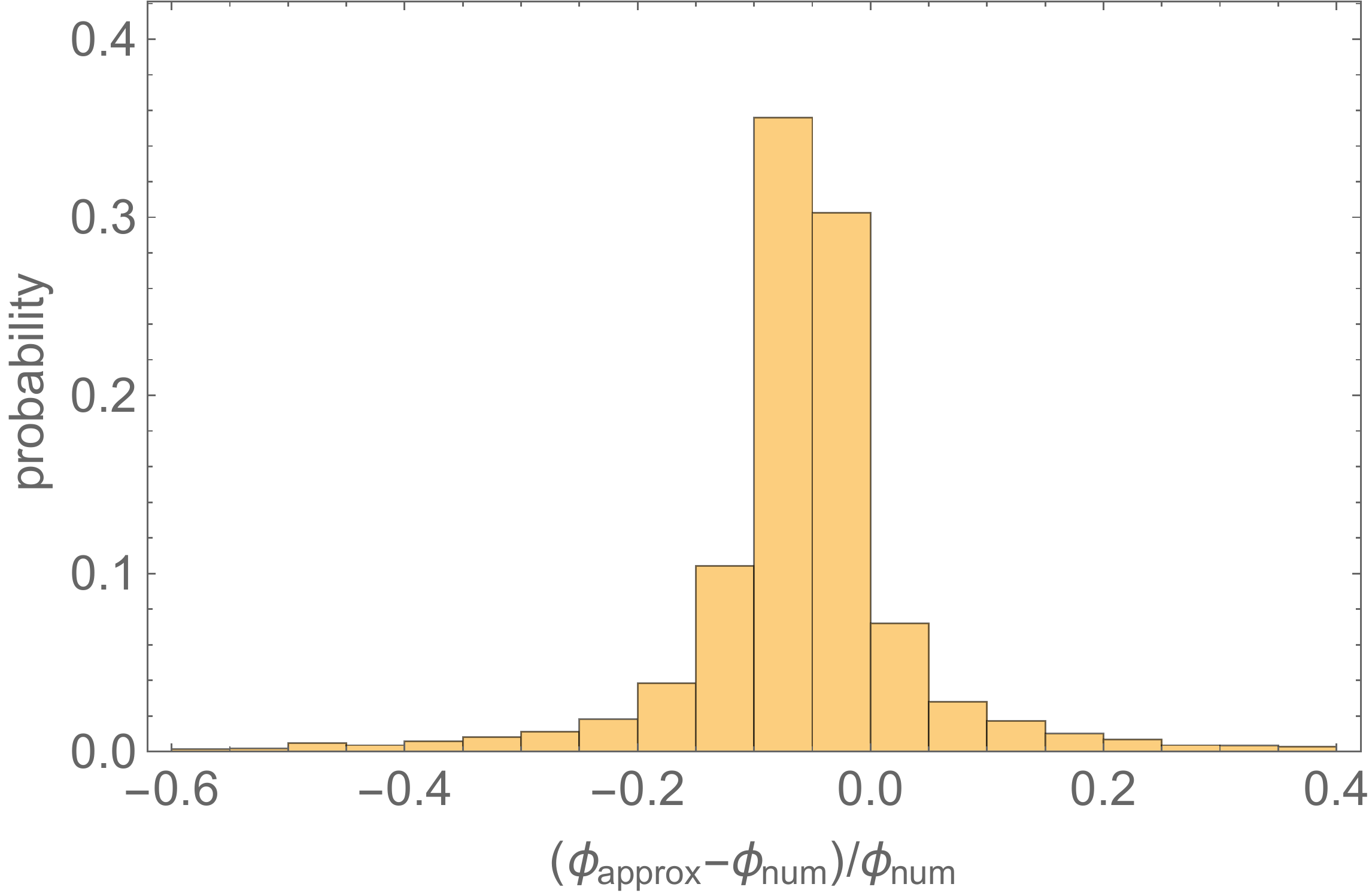}
\caption{Left: comparison of the numerical (red dots) and approximate (blue line) calculations of the heterodyne-signal phase in Fig.\ \ref{Fig-h-signal} (left). Right: histogram of $10^4$ Monte Carlo calculation of the fractional residuals of (\ref{phi}) when the wavefront misalignment is $\alpha=300$ $\mu$rad. The detector radius is $r_0=1.1$ mm, the normalized radius of the interference pattern is $w/r_0=1$. The $\we(x,y)$ flatness is constrained to $\lambda/20$.}\label{Fig-accuracy}
\end{figure}

Figure \ref{Fig-phase-profile} shows a randomly generated phase profile, $\we(x,y)$, of the interference pattern. The phase and noise of the heterodyne signal, calculated according to (\ref{phi}) and (\ref{noise-var}), are shown in Fig.\ \ref{Fig-h-signal} {\it vs.} the mean wavefront-misalignments, $\alpha_{0x}$ and $\alpha_{0y}$. To test the accuracy of the approximations made, Fig.\ \ref{Fig-accuracy} (left) compares (\ref{phi}) against the numerical integration of (\ref{phase}). As expected, the approximation looks good up to about 150 $\mu$rad wavefront misalignments; larger misalignments would require that additional terms are taken into account. In order to quantify the approximation error, we calculated the fractional residuals of $10^4$ approximated values when the wavefront misalignment is $\alpha=300$ $\mu$rad. Figure \ref{Fig-accuracy} (right) shows that, on the average, (\ref{phi}) underestimates the phase by 5\%, with a standard deviation of 10\%.

\begin{figure}
\includegraphics[width=6.5cm]{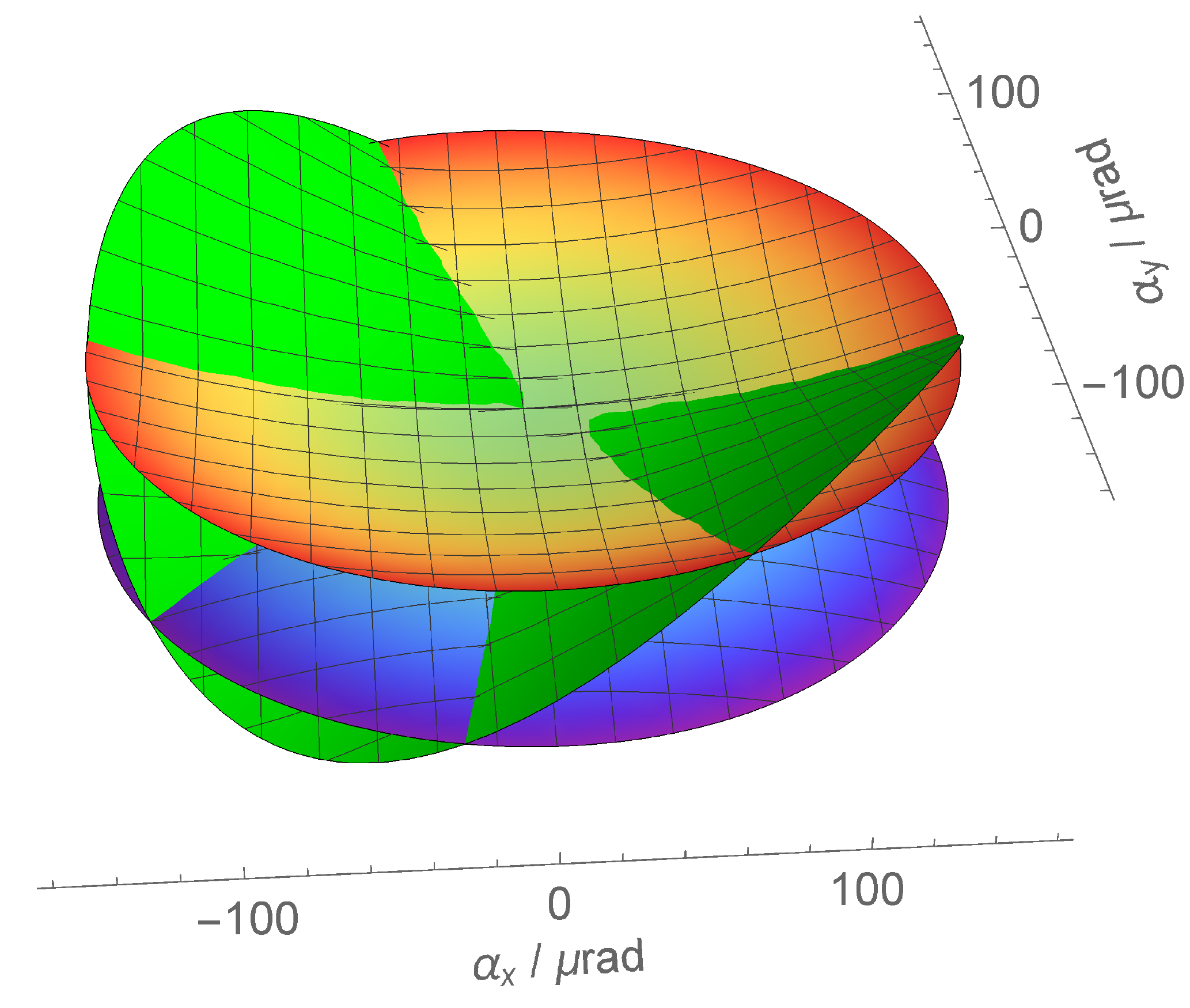}
\caption{Upper and lower bounds (standard deviations) of the heterodyne-signal phase -- see equation (\ref{phi}) -- calculated for $10^4$ random phase profiles $\we(x,y)$ constrained to a $\lambda/20$ flatness. The colours range from $-0.8$ nm (violet) to $+0.8$ nm (red). $\alpha_x$ and $\alpha_y$ are the wavefront misalignments. The detector radius is $r_0=1.1$ mm, the normalized radius of the interference pattern is $w/r_0=1$. The phases profile of Fig.\ \ref{Fig-h-signal} (left) is also shown (green).}\label{Fig-phase-bounds}
\end{figure}

When the flatness of $\we(x,y)$ is constrained to within $\lambda/20$, the phase of the heterodyne signal is bounded as shown in Fig.\ \ref{Fig-phase-bounds}. We subtracted the offset occurring when the wavefronts are aligned to make evident the dependence of the bound on the misalignment. Asymptotically, the phase standard-deviation increases with the misalignment by about $5.3$ pm/$\mu$rad.

\begin{figure}
\includegraphics[width=6.5cm]{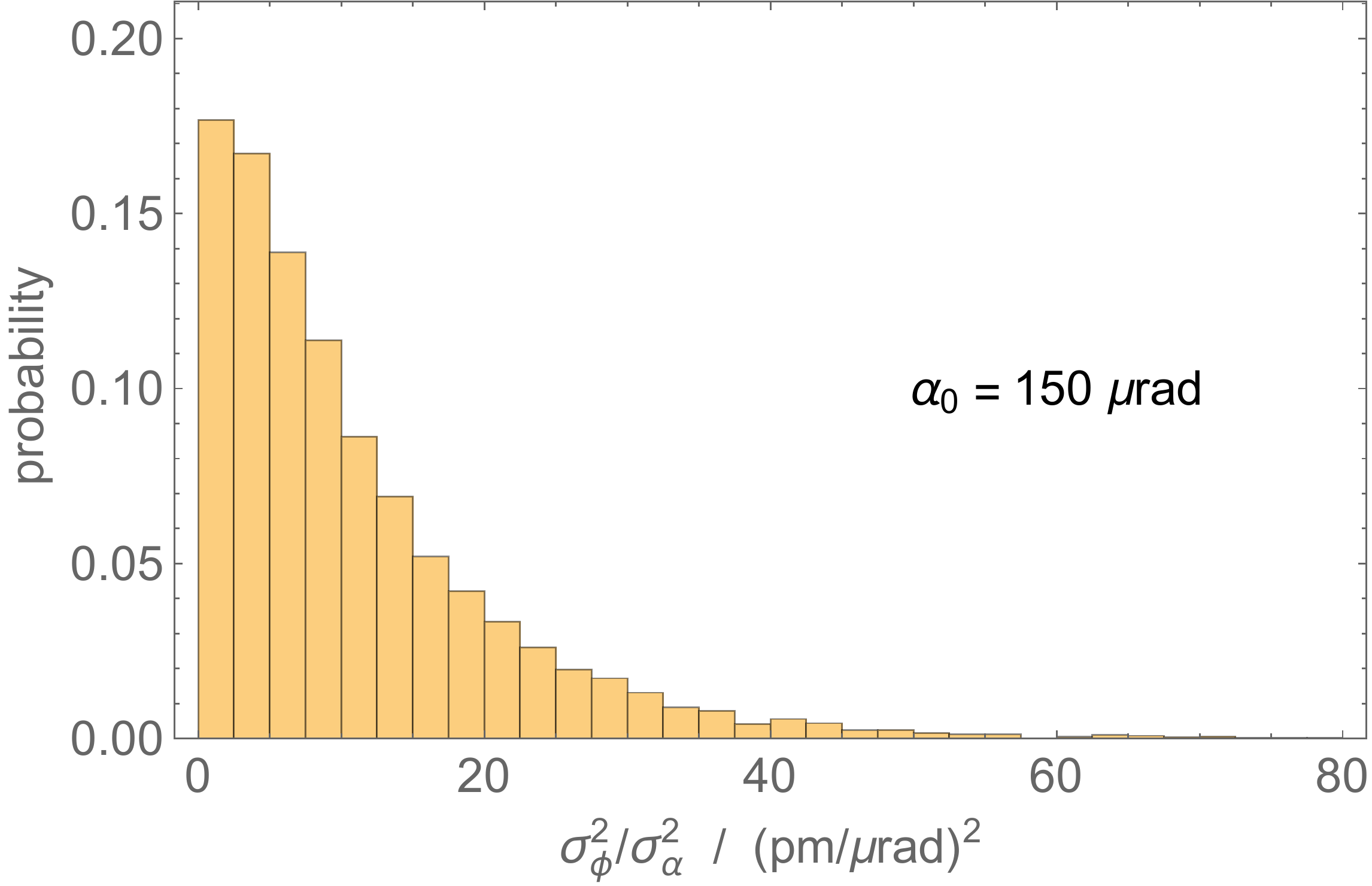}
\caption{Histogram of $10^4$ Monte Carlo calculations of the $\sigma_\phi^2/\sigma_\alpha^2$ ratio for $150$ $\mu$rad wavefront misalignment. The jitter is isotropic and has $\sigma_\alpha^2$ variance. The detector radius is $r_0=1.1$ mm, the normalized radius of the interference pattern is $w/r_0=1$. The phase profiles $\we(x,y)$ are constrained to a $\lambda/20$ flatness.}\label{Fig-phase-noise-histo}
\vspace{2mm}
\includegraphics[width=6.5cm]{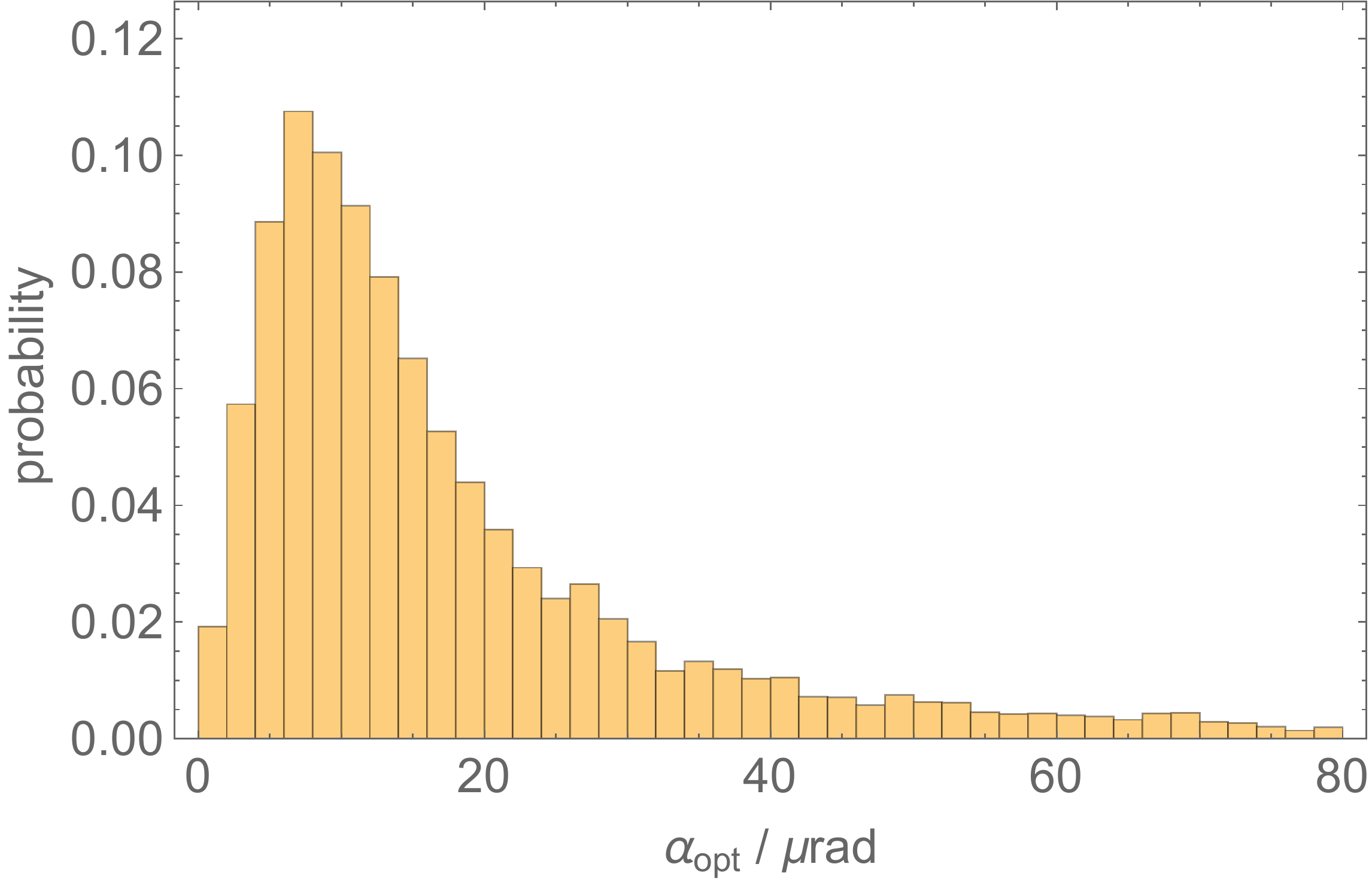}
\caption{Histogram of $10^4$ Monte Carlo calculations of the wavefronts misalignment $\alpha_{\rm opt}$ that minimize the phase noise of the heterodyne signal. The $\alpha_{\rm opt}$ distribution is uniform in the $[0, 2\pi]$ interval. The detector radius is $r_0=1.1$ mm, the normalized radius of the interference pattern is $w/r_0=1$. The phase profiles $\we(x,y)$ are constrained to a $\lambda/20$ flatness.}\label{Fig-optimal-tilt-histo}
\vspace{2mm}
\includegraphics[width=6.5cm]{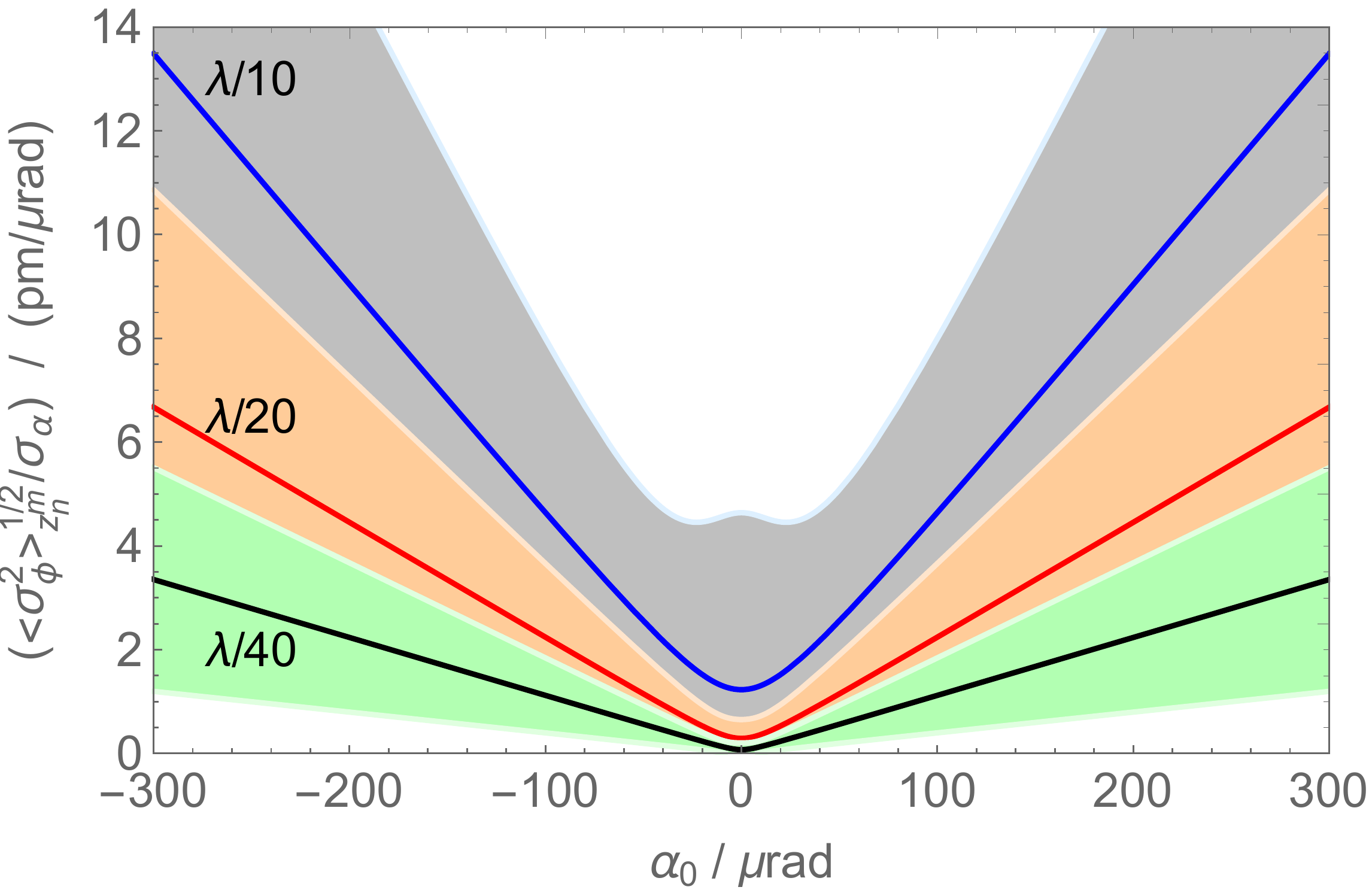}
\caption{Mean (on the indicated aberrations flatness) of $10^4$ Monte Carlo calculations of the sensitivity of the heterodyne-signal phase to the alignment jitter. The shadows indicate the $1\sigma$ confidence regions. $\alpha_0$ is the misalignment of the interfering wavefronts, The jitter is isotropic and has $\sigma_\alpha^2$ variance. The detector radius is $r_0=1.1$ mm, the normalized radius of the interference pattern is $w/r_0=1$.}\label{Fig-mean-phase-noise}
\end{figure}

In each Monte Carlo run, we calculated the ratio $\sigma_\phi^2/\sigma_\alpha^2$ between the signal and isotropic-jitter variances by using (\ref{noise-var}), where $\sigma_\alpha^2$ is the jitter variance. Figure \ref{Fig-phase-noise-histo} shows the distribution of the results in the case of a $\alpha_0=150$ $\mu$rad wavefront misalignment.

The noise is minimum when the wavefront jitters about the stationary point of (\ref{phi}). {\colr Hence, by solving $\nabla \phi(\zeta_x,\zeta_y)=0$, the optimal tilt aberrations are}
\numparts\begin{eqnarray}\label{a-offset}
 \zeta_x^{\rm opt} &= \frac{b_{01}b_{11}-2b_{10}b_{02}}{4b_{02}b_{20}-b_{11}^2} , \\
 \zeta_x^{\rm opt} &= \frac{b_{10}b_{11}-2b_{01}b_{20}}{4b_{02}b_{20}-b_{11}^2} .
\end{eqnarray}\endnumparts
Figure \ref{Fig-optimal-tilt-histo} shows the distribution of the optimal angle $\alpha_{\rm opt}=|\alpha_x^{\rm opt}, \alpha_y^{\rm opt}|$, where the optimal misalignments are linked to $\zeta_x^{\rm opt}$ and $\zeta_y^{\rm opt}$ by (\ref{hvt}-$b$). The most probable optimizing angle is about a misalignment of 7 $\mu$rad. However, the distribution has a long tail, not represented in figure \ref{Fig-optimal-tilt-histo}.

{\colr Providing that the jitter is isotropic and according to (\ref{mean-phi}), the sensitivity of the signal-phase to the alignment jitter -- averaged over arbitrary $\we(x,y)$ aberrations -- is
\begin{equation}
 \frac{ \langle \sigma_\phi^2 \rangle_{z_n^m}^{1/2}}{\sigma_\alpha} = \sqrt{g_0 + g_2\alpha_0^2} .
\end{equation}
It was obtained by fitting (\ref{mean-phi}) to the Monte Carlo average of (\ref{noise-var}); the result is shown in Fig.\ \ref{Fig-mean-phase-noise}. Table \ref{table:1} shows the best-fit values of the $g_0$ and $g_2$ parameters. The standard deviation of $\sigma_\phi/\sigma_\alpha$ was approximated by
\begin{equation}
 \frac{{\rm std}(\sigma_\phi)_{z_n^m}}{\sigma_\alpha} \approx
 \frac{{\rm std}(\sigma_\phi^2)_{z_n^m}}{2 \sigma_\alpha \langle \sigma_\phi^2 \rangle_{z_n^m}^{1/2}} ,
\end{equation}
where ${\rm std}(\square)_{z_n^m}$ is the standard deviation of $\square$ calculated with respect the joint distribution of $z_n^m$ and ${\rm std}(\sigma_\phi^2)_{z_n^m}$ was obtained from the Monte Carlo standard-deviation of (\ref{noise-var}).}

\begin{table}
\caption{\label{table:1} Model parameters of the average sensitivities of the heterodyne-signal phase to the alignment jitter of the interfering wavefronts, see (\ref{mean-phi}). The means are calculated over phase profiles of the interference pattern constrained to $\lambda/10$, $\lambda/20$, and $\lambda/40$.}
\begin{indented}
\item[]\begin{tabular}{lccc}
\br
 &$\lambda/10$ &$\lambda/20$ &$\lambda/40$ \\
\mr
$g_0\,/\,({\rm pm}/\mu{\rm rad})^2$                &1.5 &$8.5\times 10^{-2}$ &$5.4\times 10^{-4}$ \\
$g_2\,/\,({\rm pm}/\mu{\rm rad})^2/\mu{\rm rad}^2$ &$2.0\times10^{-3}$ &$4.9\times 10^{-4}$ &$1.2\times 10^{-4}$ \\
\br
\end{tabular}
\end{indented}
\end{table}

To establish a criterion for the quality of the interfering wavefronts, we repeated the Monte Carlo calculation in the case of arbitrary $\lambda/10$ and $\lambda/40$ aberrations. The asymptotic average sensitivity is
\begin{equation}
 \langle \sigma_\phi^2 \rangle_{z_n^m}^{1/2} \approx \sqrt{g_2}\alpha_0 \sigma_\alpha ,
\end{equation}
where $\sqrt{g_2}$ increases from $1.1\times 10^{-2}$ pm/$\mu$rad$^2$ to $4.5\times 10^{-2}$ pm/$\mu$rad$^2$ when the quality of the interfering wavefront decreases from $\lambda/40$ to $\lambda/10$.

\section{Conclusions}
Heterodyne interferometry, where laser beams are simultaneously transmitted and received by onboard telescopes, monitors the separation of the LISA's spacecraft down to picometre sensitivity. The telescopes' pointing is continuously corrected to compensate for the disturbances \cite{Dong:2014}, but the feedback loop jitters the propagation directions of the transmitted and received beams.

Due to the receiver tilts and jitter, the interfering wavefronts are misaligned and jittered by angles scaled-up by the telescope magnification. If the wavefronts of the two interfering beams match, the jitter does not affect the phase of the heterodyne signal. However, wavefront aberrations couple to the jitter and induce a noise.

{\colr The equations (\ref{phi}) and (\ref{noise-var}) give the phase and phase noise of the heterodyne signal in terms of the radius $w$ of the interference pattern and the lowest-order Zernike aberrations of the phase profile. On these bases, we carried out a Monte Carlo calculation of the jitter-induced noise for Gaussian intensity profiles of the interfering beams and arbitrary wavefront misalignments and aberrations. Eventually, we estimated the phase sensitivity to isotropic jitter for $r_0=w$ detector radius, arbitrary $\lambda/10$, $\lambda/20$, and $\lambda/40$ aberrations, and up to 300 $\mu$rad wavefront misalignments. The average sensitivity is always less than the required 25 pm/$\mu$rad value \cite{Chwalla:2016}. However, owing to the large dispersion, a wavefront quality at least equal to $\lambda/20$ might be necessary. These results extend and complement our investigation of how the measured spacecraft distance is coupled to the transmitter jitter \cite{Sasso:2018} and open the way to a full start-to-end analysis of the phase noise.}

The assumption that aberrations other than defocus, astigmatism, coma, trefoil and spherical are negligible might be optimistic, and future work must examine the impact of higher-frequency aberrations. The experimental observation reported in \cite{Balsamo:2003,Sasso:2016} suggests that the Zernike spectra of the interfering wavefronts might have high-frequency components originated in the beam path through the optical bench and the receiving telescope.

\section{Acknowledgments}
This work was funded by the European Space Agency (contract 1550005721, Metrology Telescope Design for a Gravitational Wave Observatory Mission).

\appendix
\setcounter{section}{1}
\section*{Appendix}
\begin{figure}\centering
\includegraphics[width=7.5cm]{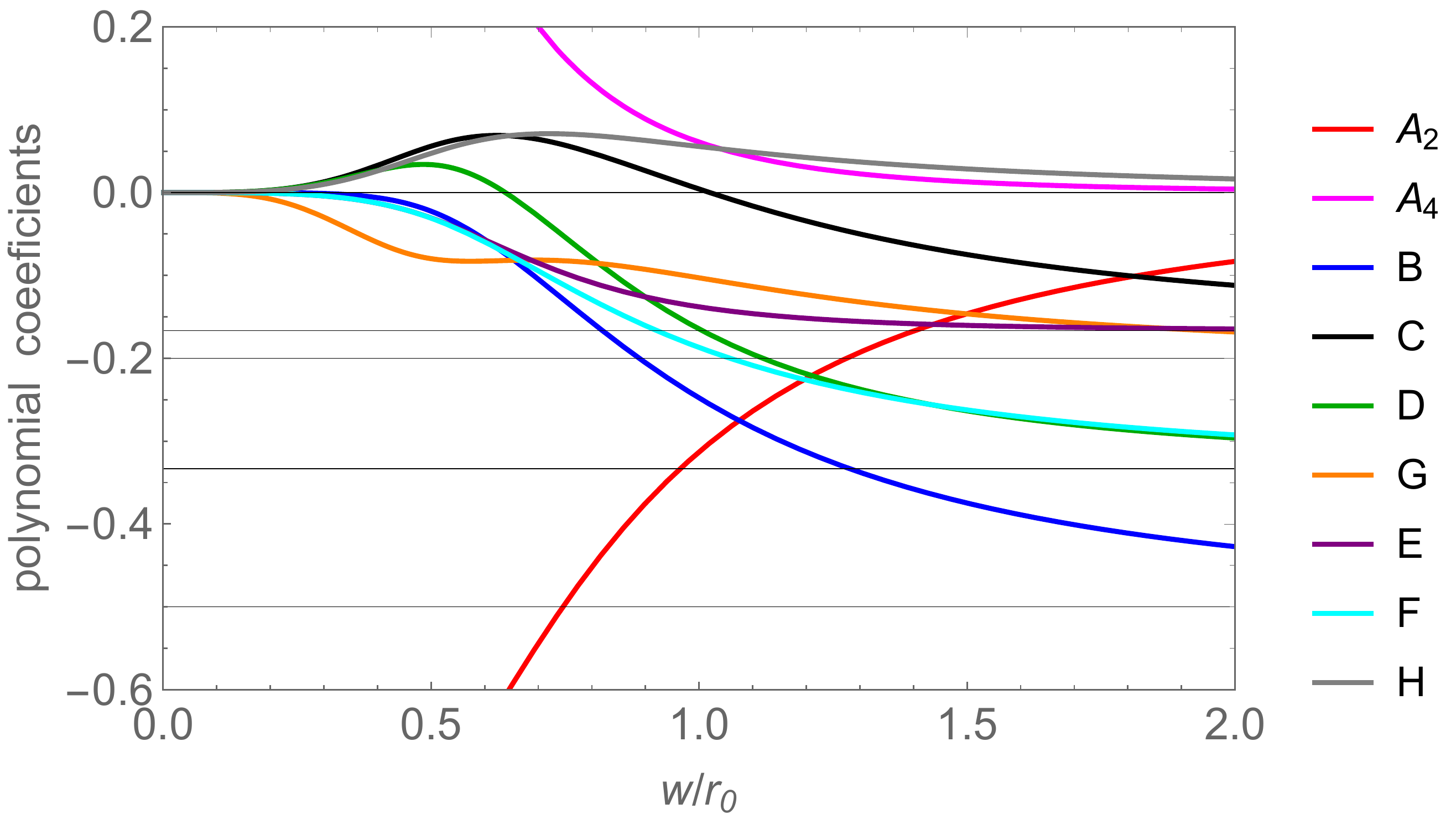}
\caption{Coefficients of the (\ref{phi}) polynomials {\it vs.} the $w/r_0$ ratio. The horizontal lines are the asymptotic values (flat intensity-profile). The limits of $A_2$ and $A_4$ when $w/r_0 \rightarrow 0$ are $\pm 1$.} \label{abcde}
\end{figure}

The coefficients of the extra phase of the heterodyne signal (\ref{phi}) are
\numparts\begin{eqnarray}\fl\label{bxy}
 b_{00} &=  A_2 z_2^0 + A_4 z_4^0, \\ \fl
 b_{10} &= B\cos(\theta_3^3-\theta_2^2)|z_3^3||z_2^2| + C\cos(\theta_2^2-\theta_3^1)|z_3^1||z_2^2| \\ \fl \nonumber
        &+ D\cos(\theta_3^1)|z_3^1|z_2^0 + G\cos(\theta_3^1)|z_3^1|z_4^0 ,\\ \fl
 b_{01} &= B\sin(\theta_3^3-\theta_2^2)|z_3^3||z_2^2| + C\sin(\theta_2^2-\theta_3^1)|z_3^1||z_2^2| \\ \fl \nonumber
        &+ D\sin(\theta_3^1)|z_3^1|z_2^0 + G\sin(\theta_3^1)|z_3^1|z_4^0 ,\\ \fl
 b_{20} &= Ez_2^0 + F\cos(\theta_2^2)|z_2^2| + Hz_4^0 ,\\ \fl
 b_{02} &= Ez_2^0 - F\cos(\theta_2^2)|z_2^2| + Hz_4^0,\\ \fl
 b_{11} &= 2F\sin(\theta_2^2)|z_2^2| ,
\end{eqnarray}\endnumparts
where, by measuring the 1/e$^2$ radius of the interference in terms of the detector radius,
\numparts\begin{eqnarray}\fl\label{Fterm}
 A_2 &=  \frac{ 1 + e^{2/\tw^2}}{1-e^{2/\tw^2}} + w'^2 , \\ \fl
 A_4 &=  1 + \frac{3(1+e^{2/\tw^2})\tw^2}{1-e^{2/\tw^2}}+ 3w'^2, \\ \fl
 B &= -\frac{2 + 3 \tw^2 + 3 \tw^4}{1 - e^{2/\tw^2}} - \frac{3}{2}\tw^6 ,\\ \fl
 C &= -\frac{2 + 5 \tw^2 + (7 + 2 e^{2/\tw^2}) \tw^4}{1 - e^{2/\tw^2}} - \frac{9}{2}\tw^6 ,\\ \fl
 D &=  \frac{4e^{2/\tw^2} + 12e^{2/\tw^2}\tw^2 - 2(2+e^{2/\tw^2})(1 - e^{2/\tw^2})\tw^4}
       {(1 - e^{2/\tw^2})^2} - 6w'^2 ,\\ \fl \nonumber
 G &=  \frac{12e^{2/\tw^2}\tw^2 - 6(2-9e^{2/\tw^2}\!+e^{4/\tw^2})\tw^4 - 6(7-2e^{2/\tw^2}\!+5e^{4/\tw^2})\tw^6}
       {(1 - e^{2/\tw^2})^2} - 45\tw^8, \\ \fl
   & \\ \fl
 E &=  \frac{2e^{2/\tw^2}}{(1 - e^{2/\tw^2})^2} - \frac{1}{2}w'^4 ,\\ \fl
 F &= -\frac{ 1 + \tw^2 }{1 - e^{2/\tw^2}} - \frac{1}{2}w'^4 , \\ \fl
 H &= \frac{6e^{2/\tw^2}\tw^2 - 3(1-e^{4/\tw^2})\tw^4/2}{(1 - e^{2/\tw^2})^2} - w'^6 .
\end{eqnarray}\endnumparts
Figure\ \ref{abcde} shows the $A_2,\, A_4,\, B,\, ...\, H$ coefficients {\it vs.} the normalized radius $\tw=w/r_0$. Apart from the $\tw\rightarrow\tw/\sqrt{2}$ transformation, (\ref{Fterm}-$i$) are the same as given in \cite{Sasso:2018}.

\section*{References}
\bibliography{LISA near field_ArXiv}

\end{document}